\documentclass[aps,prb,twocolumn,showpacs]{revtex4}

\usepackage{pstricks,pst-plot}           
\usepackage{latexsym}
\usepackage{bbm,epsf,graphicx,graphics,epsfig,psfrag}
\usepackage{mathrsfs,amsmath,amssymb}

\def\r{\varrho}

\newcommand{\pf}[1]{\frac{\partial}{\partial #1}}

\newcommand{\pfrq}[1]{\partial^{2}_{ #1}}

\begin{document}

\title{Effect of spin-orbit coupling on zero-conductance resonances in
  asymmetrically coupled one-dimensional rings} 

\author{Urs Aeberhard}
\affiliation{Condensed Matter Theory, Paul Scherrer Institute, CH-5232 Villigen, Switzerland}
\author{Katsunori Wakabayashi}
\affiliation{Theoretische Physik, ETH-H\"onggerberg, CH-8093 Z\"urich, Switzerland}
\author{Manfred Sigrist}
\affiliation{Theoretische Physik, ETH-H\"onggerberg, CH-8093 Z\"urich, Switzerland}
\begin{abstract}

The influence of Rashba spin-orbit coupling on zero conductance
resonances appearing in one-dimensional conducting rings
asymmetrically coupled to two leads is investigated. For this purpose,
the transmission function of the corresponding one-electron scattering
problem is derived analytically and analyzed in the complex energy
plane with focus on the zero-pole structure characteristic of transmission
(anti)resonances. The lifting of real conductance zeros due to
spin-orbit coupling in the asymmetric Aharonov-Casher ring is
related to the breaking of spin reversal symmetry 
in analogy to the time-reversal
symmetry breaking in the asymmetric Aharonov-Bohm ring.  
\end{abstract} 


\maketitle

\section{Introduction\label{sec:1}}

An important feature of one-dimensional ring shaped conductors or electronic
devices is the appearance of quantum interference effects under the influence of
electromagnetic potentials, known as Aharonov-Bohm\cite{ab:59} (AB) and
Aharonov-Casher\cite{ac:84} (AC) effect. In numerous investigations, the
transmission properties of mesoscopic AB and AC-rings coupled to
current leads were studied under various aspects such as AB-flux and
coupling dependence of resonances\cite{buettiker:84}, geometric
(Berry) phases\cite{loss:90,ady:92,alg:93,quian:94,hentschel:03} and spin flip,
precession and interference effects\cite{yi:97,nitta:99,hentschel2:03,
  frustaglia:04,molnar:04}. Most of the investigated models use
symmetrically coupled rings. There are however mesoscopic systems like
nanographite ribbons showing conductance properties that are based on 
asymmetric configurations \cite{wakabayashi:01},
giving rise to a specific dip structure of anti-resonances (zero-conductance
resonances) in the model transmission. The effects of asymmetry on the
transmission were considered mainly in quantum network
models\cite{wu:91,yi:03,bercioux:04}. In quasi 1d systems, real conductance zeros
appear under the condition of conserved 
time reversal symmetry\cite{lee:99,lee:01} (TRS). The
(anti)resonances in the transmission due to local quasi-bound states
correspond to a specific zero-pole structure in the complex energy
plane\cite{porod:93,porod:94,deo:94,price:93}. The application of an
external magnetic field modifies this zero-pole structure, shifting
the transmission zeros away from the real axis, with the shift as a
function of the AB-phase\cite{kim:02}. Thus, the lifting of
conductance zeros is related to the breaking of TRS. 

In this paper, the influence of spin-orbit coupling (SOC) on
zero-conductance resonances in asymmetrically coupled rings is
investigated by means of an AC-ring where an effective in-plane magnetic field
results from the \emph{Rashba} effect\cite{rashba:60} of moving
electrons in the presence of an electric field perpendicular to the
ring plane, as considered in
Ref. \onlinecite{frustaglia:04} and \onlinecite{molnar:04}. This means that the role of
time reversal symmetry is now transfered to inversion symmetry
(parity). We will show that parity connected with the 
Rashba spin orbit coupling can be viewed in an analogous way as
the case of time reversal symmetry for spinless particles. 

This paper is organized as follows. In Sec. \ref{sec:2}, a
single-particle description of the one-dimensional ring subject to
Rashba-SOC in terms of Hamiltonian, eigenstates and eigenenergies is
given, following Ref. \onlinecite{molnar:04,frustaglia:04}. The section concludes with the
results for the transmission of the asymmetric AC-ring in the one-electron scattering
picture which is derived in the appendix. The analytic expression for the
transmission function is analyzed in Sec. \ref{sec:3} with focus on
geometry and SOC dependence of the transmission
zeros. Sec. \ref{sec:4} contains a symmetry argument which establishes
an analogy between formation  and lifting of the zeros due to
Rashba-SOC in the AC-ring and the corresponding effects on spinless
electrons due to the magnetic field in the AB-ring. The main results
are summarized in the conclusions of Sec. \ref{sec:5}. 

\section{AC-ring in single particle picture \label{sec:2}}
The coupling of electron spin and orbital degrees of freedom is due to
the magnetic field generated in the reference frame of a moving
electron by an electric field in the reference frame of the
laboratory. In two dimensional systems (e.g. due to the presence of a
confinement potential along a specific direction), an important
contribution of electric fields is the Rashba effect, a consequence of
lack of inversion symmetry, that causes a spin band-splitting
proportional to the momentum. In the ring system under consideration,
the Rashba field results from the asymmetric confinement along the
direction perpendicular to the ring plane. 
\subsection{Hamiltonian}
In the following investigation of one-dimensional rings, $z$ is chosen as the direction of
confinement, perpendicular to the plane of motion. The various
SO-coupling mechanisms are accounted for using the following model
Hamiltonian: 
\begin{equation}
\hat{H}_{SO}=\frac{\alpha}{\hbar}(\hat{\vec{\sigma}}\times\hat{\vec{p}})_{z}=i\alpha\left(\hat{\sigma}_{y}\pf{x}-\hat{\sigma}_{x}\pf{y}\right), 
\end{equation}
where $\frac{\hbar}{2}\hat{\vec{\sigma}}$ is the spin
operator in terms of the Pauli spin matrices, $\hat{\vec{\sigma}}=(\sigma_{x},\sigma_{y},\sigma_{z})$ and $\alpha$ is the Rashba parameter characterizing the strength
of the SOC corresponding to an electric field $\vec E_{R}=\left(0,0,E_{z}\right)$ in $z$-direction, arising from a
potential $V(z)$ due to structural or confinement asymmetry. In polar
coordinates $x=r\cos\varphi$ and $y=r\sin\varphi$ the total
Hamiltonian in effective mass approximation reads\cite{meijer:02} 
\begin{align}
\hat{H}(r,\varphi)=&-\frac{\hbar^{2}}{2m^{*}}\left[\pfrq{r}+\frac{1}{r}\pf{
    r}+\frac{1}{r^{2}}\pfrq{\varphi}\right]-\frac{i\alpha}{r}(\cos\varphi\sigma_{x}\nonumber\\&+\sin\varphi\sigma_{y})\pf{\varphi}+i\alpha(\cos\varphi\sigma_{y}-\sin\varphi\sigma_{x})\pf{r},  
\end{align}
with the effective mass $m^{*}$.
In the case of a one-dimensional ring, a confining potential
$V(r)$ needs to be added in order to force the electron wave
functions to be localized on the ring in the radial
direction. It is shown in Ref. \onlinecite{meijer:02} that the exact
form of the confining potential is not essential. A simple
possibility is the harmonic potential centered around $r=\r$,
$V(r)=\frac{1}{2}K(r-\r)^{2}$ where $ \r $ is the radius of the ring. 
\begin{figure}[h!]
    \begin{center}
    \footnotesize
    \psfrag{a}{$(a)$}
    \psfrag{b}{$(b)$}
    \psfrag{Br}{$\vec{B}_{R}$}
    \psfrag{up}{$|\uparrow\rangle$}
    \psfrag{down}{$|\downarrow\rangle$}
    \psfrag{phi}{$\varphi$}
    \psfrag{E}{$\vec{E}_{R}$}
    \psfrag{z}{$\vec{E}_{R}$}
    \psfrag{p}{+}
    \psfrag{m}{-}
    \psfrag{g}{$\tilde{\theta}$}
    \psfrag{k}{$\vec{v}_{g}$}
    \psfrag{x}{x}
    \psfrag{xs}{x'}
    \psfrag{phis}{$\varphi'$}
    \psfrag{I}{I}
    \psfrag{II}{II}
    \psfrag{r}{$\r$} 
    \psfrag{A}{A}
    \psfrag{B}{B}
    \includegraphics[width=2.5in]{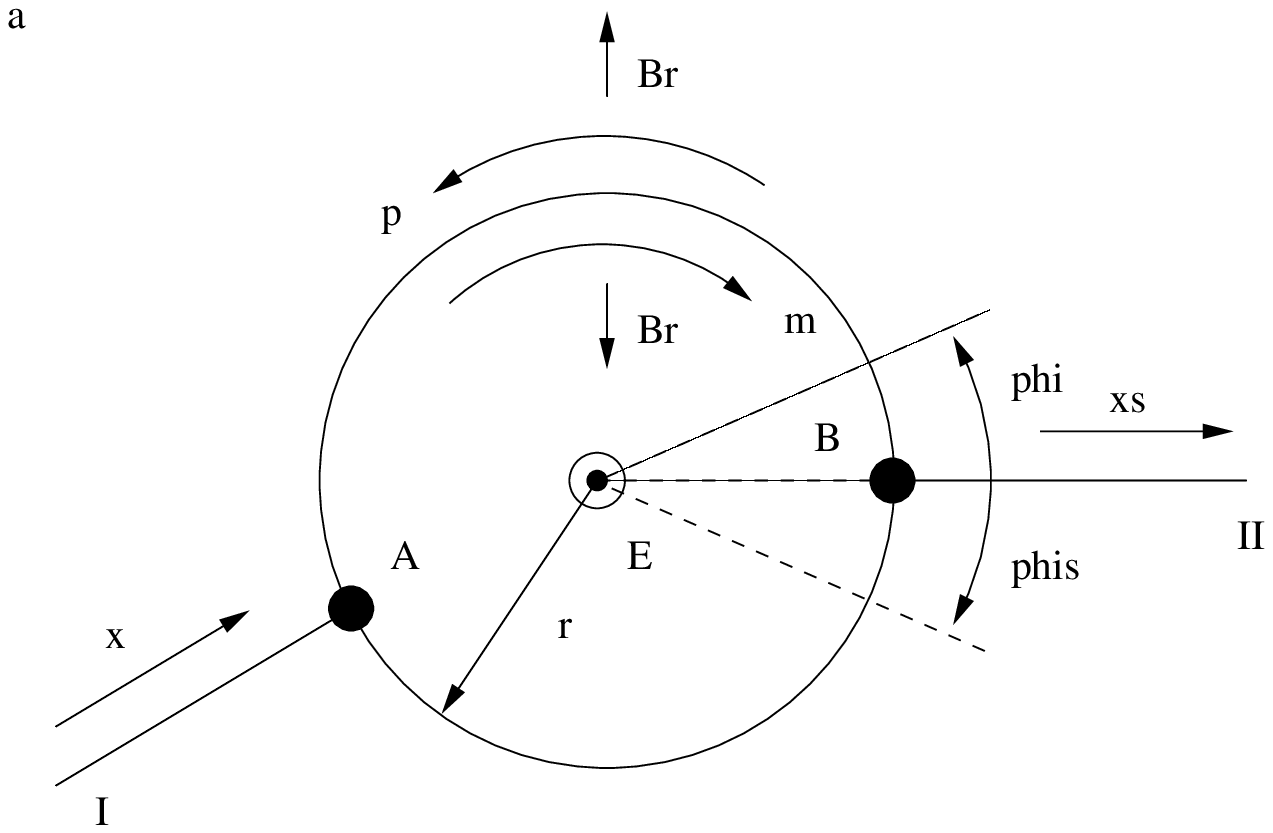}\vspace{0.5cm}\newline
\includegraphics[width=2.6in]{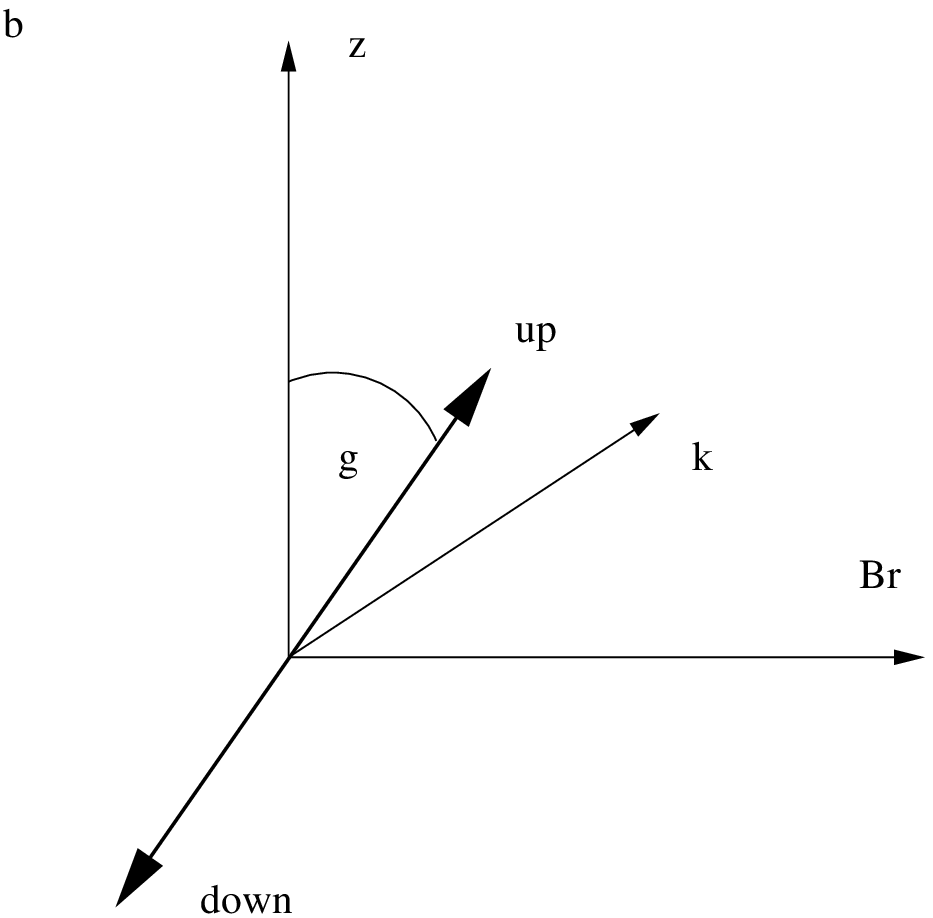} 
    \end{center}
    \caption{(a) Momentum dependent in-plane Rashba field
    $\vec{B}_{R}$, (b) Up and down spin eigenstates do not generally
    align with the Rashba field $\vec{B}_{R}$, but make a tilt angle
    $\theta$ with the electric field $\vec{E}_{R}$ perpendicular to the
    ring plane ($\vec{E}_{R}$, $\vec{B}_{R}$ and $\vec{v}_{g}$ form an orthogonal coordinate system). } 
    \label{fig:tiltangle1}
\end{figure}
Considering only the lowest radial mode,
the resulting one-dimensional Hamiltonian for fixed radius $\r$ is
(see Ref. \onlinecite{meijer:02} for a complete derivation) 
\begin{align}
\hat{H}_{1D}(\varphi)=&\langle R_{0}(r)|\hat{H}(r,\varphi)|R_{0}(r)\rangle\nonumber\\
=&-\frac{\hbar^2}{2m^{*}\r^{2}}\frac{\partial^{2}}{\partial\varphi^{2}}-\frac{i\alpha}{\r}(\cos\varphi\sigma_{x}\nonumber\\&+\sin\varphi\sigma_{y})\frac{\partial}{\partial\varphi}-\frac{i\alpha}{2\r}(\cos\varphi\sigma_{y}-\sin\varphi\sigma_{x}).
\label{eq:1dham} 
\end{align}
The last term in the above expression for the 1D-Hamiltonian encodes
the correction due to the radial confinement. The Hamiltonian in Eq.(\ref{eq:1dham}) can be written in a dimensionless
form\cite{molnar:04}, 
\begin{equation}
 H = \frac{2m^{*}\r^{2}}{\hbar^{2}}\hat
 H_{1D}=\left(-i\pf{\varphi}+\frac{\beta}{2}\sigma_{r}\right)^{2}\label{eq:Hamiltonian} 
\end{equation}
where $\beta=2\alpha m^{*}\r/\hbar^{2}$ is the
dimensionless SOC-constant,
$\sigma_{r}=\cos\varphi\sigma_{x}+\sin\varphi\sigma_{y}$ and
the additive constant $-\beta^{2}/4$ was
neglected\footnote{The neglected
term in \eqref{eq:Hamiltonian} produces a spin independent modification of the spectrum,
which should not influence the interference effects.}. 
\newline
\subsection{Eigenstates and energy spectrum}
The eigenstates of Hamiltonian \eqref{eq:Hamiltonian} follow as the solution of
the time-independent Schr\"odinger equation and have the general form
\cite{frustaglia:04,molnar:04} 
\begin{equation}
\Psi_{n}^{\sigma}(\varphi)=e^{i n\varphi}\chi^{\sigma}(\varphi),
\label{eq:ringeigstate}
\end{equation}
where $n$ is the orbital quantum number and
$\sigma=\uparrow,\downarrow\cong\pm1$ labels the spin. For the
isolated ring, $n\in\mathbbm{Z}$, but when coupled to leads, $n$ can adopt any real number allowed by energy, depending
on spin and direction of motion. 
The spinors
$\chi^{\sigma}(\varphi)$ are generally not aligned with the momentum
dependent and spatially varying
Rashba-field $\vec{B}_{R}(r)=2\beta(\hat z\times \vec
p)/\varrho$, but make a tilt angle $\tilde\theta=\pi/2-\theta$ given by $\tan\theta=-\beta$
relative to the direction of the electric field $\vec{E}_{R}$ (see
Fig. \ref{fig:tiltangle1}). 
The energy eigenvalues of the states in Eq.(\ref{eq:ringeigstate}) are\cite{molnar:04} 
\begin{equation}
E_{n}^{\sigma}=\left(n-\Phi_{AC}^{\sigma}/2\pi\right)^{2}.\label{eq:eigenenergy}
\end{equation}
with the Aharonov-Casher phase\cite{frustaglia:04}
\begin{equation}
\Phi_{AC}^{\sigma}=-\pi\left(1-\sigma\sqrt{\beta^{2}+1}\right).
\end{equation}
At fixed energy $E$, the dispersion relation yields the quantum numbers $n_{\lambda}^{\sigma}(E)$ through
\begin{equation}
n_{\lambda}^{\sigma}(E)=\lambda \sqrt{E} +\Phi_{AC}^{\sigma}/2\pi,\quad \lambda=\pm.
\end{equation}
For a plane wave arriving from lead I with wave vector $k$ we get
\begin{equation}
n_{\lambda}^{\sigma}(k)=\lambda k\r+\Phi_{AC}^{\sigma}/2\pi.
\label{eq:n}
\end{equation}
The sense of propagation is determined by the sign of the group
velocity, which in the latter case is given by 
\begin{equation}
v_{g,\lambda}^{\sigma}=\frac{\hbar}{2m^{*}\r}\frac{dE_{n_{\lambda}^{\sigma}}^{\sigma}}{dn_{\lambda}^{\sigma}}=\frac{\hbar}{2m^{*}\r}(n_{\lambda}^{\sigma}-\Phi_{AC}^{\sigma}/2\pi)=\lambda 
k\r, 
\end{equation}
$\lambda$ thus encoding the traveling direction. The quantum numbers
for different spin and sense of propagation are related by 
\begin{equation}
n_{\lambda}^{\sigma}=-\left(n_{-\lambda}^{-\sigma}+1\right).\label{eq:qn}\end{equation}\newline
The corresponding eigenstates of the closed ring are
\begin{align}
\Psi_{\lambda}^{\sigma}(\varphi)&=e^{in_{\lambda}^{\sigma}\varphi}
\left(\begin{array}{c}
\sin\Big(\frac{\theta}{2}+\frac{\pi}{4}(1+\sigma)\Big)\\
-\cos\Big(\frac{\theta}{2}+\frac{\pi}{4}(1+\sigma)\Big)e^{i\varphi}\\
\end{array}\right) \frac{1}{\sqrt{2\pi}},\label{eq:es}\\
\sigma&=\pm1,\quad\lambda=\pm.\nonumber
\end{align}
These eigenstates differ from the solutions of the free system by the phase factors in the spin part. 
\subsection{Current}
In order to investigate transport in our quantum mechanical system, an
expression for the probability current density is requested. The
probability current density $j$ is determined by inserting the
Schr\"odinger equation 
\begin{equation}
i\hbar\frac{\partial\Psi}{\partial t}=H\Psi,
\end{equation}
with $H$ from Eq.(\ref{eq:Hamiltonian}), and its adjoint into the
continuity equation imposed by probability conservation, 
\begin{equation}
\frac{\partial\rho}{\partial t}+\frac{\partial j}{\partial \varphi}=0,
\end{equation}
where $\rho=|\Psi|^2$ denotes the probability density. The probability
current density can
be expressed in terms of velocity operators: 
\begin{equation}
j=\frac{1}{2}\left(\Psi^{\dagger}(\hat{v}\Psi)+\Psi(\hat{v}\Psi)^{\dagger}\right).
\label{eq:current}
\end{equation}
The velocity operators are derived from the Hamiltonian by\cite{molenkamp:01}, 
\begin{equation}
\hat{v}=\frac{\partial \hat{H}}{\partial\hat{p}}
\end{equation}
where $\hat{p}$ is the momentum operator, whose explicit form depends
on the coordinate system\footnote{Note that using $H$, we are still
  working with dimensionless quantities.}: 
\begin{equation}
\hat{p}_{\varphi}=-i\pf{\varphi}~(\mathrm{ring})\quad
\mathrm{and}\quad\hat{p}_{x}=-i\r\pf{x}~(\mathrm{leads}). 
\end{equation}
In absence of SOC, only the kinetic energy term of the Hamiltonian
contributes to the velocity operators, which in this case are  
\begin{equation}
\hat{v}_{0}(\varphi)=-2i\pf{\varphi}\quad \mathrm{and}\quad
\hat{v}_{0}(x)=-2i\r\pf{x}.\label{eq:v0} 
\end{equation}
For finite SOC, $H_{SO}$ yields an additional term for the ring (assuming zero SOC in the leads)
\begin{equation}
\hat{v}_{SO}(\varphi)=\beta\sigma_{r}(\varphi)~\mathrm{and}~\hat{v'}_{SO}(\varphi')=\beta\sigma_{r}'(\varphi'),\label{eq:vso}  
\end{equation}
where $\sigma_{r}'(\varphi')\equiv\sigma_{r}(-\varphi')=\cos\varphi'\sigma_{x}-\sin\varphi'\sigma_{y}$.

The total velocity operator to consider in the expression of the
probability current density given by Eq.\eqref{eq:current} is 
\begin{equation}
\hat v=\hat{v}_{0}+\hat{v}_{SO}.\end{equation} 
The above results will be used when investigating the lead and ring-currents
in the appendix.
\subsection{Transmission amplitude from the one-electron scattering formalism}

Conductance in mesoscopic structures can be expressed by means of the Landauer
conductance formula\cite{b"uttiker:85,landauer:87}, which in our case reads
\begin{equation}
G=\frac{e^{2}}{h}\sum_{\sigma=\uparrow,\downarrow}|T_{\sigma}|^{2},\label{eq:land}
\end{equation}
where $T_{\sigma}$ is the (spin dependent) transmission
amplitude\footnote{Generally, $
G=\frac{e^{2}}{h}\sum_{\sigma,\sigma'=\uparrow,\downarrow}|T_{\sigma\sigma'}|^{2}$,
where $T_{\sigma\sigma'},~\sigma\neq\sigma'$ is the spin flip amplitude. The
conductance does not depend on the choice of spinor basis, i.e. it is invariant
under spin rotation, we can therefore make use of the ring-spinor basis where
the spin flip amplitudes vanish, and define $T_{\sigma\sigma}\equiv
T_{\sigma}$.}. The
previously obtained expressions for wavefunction and current are now used to
calculate the transmission amplitude for the ring system from the proper
requirements on wave function continuity and probability current conservation\cite{griffiths:53}. The calculation is
performed in the appendix and follows Ref. \onlinecite{molnar:04} and
\onlinecite{frustaglia:04}. It yields the transmission amplitude 
\begin{widetext}
\begin{equation}
T_{\sigma}(\phi,\beta,\gamma)=\frac{4i\Big[e^{i\frac{\Phi_{AC}^{\sigma}}{2}(1-\gamma)}\sin\big(\frac{\phi}{2}(1+\gamma)\big)+e^{-i\frac{\Phi_{AC}^{\sigma}}{2}(1+\gamma)}\sin\big(\frac{\phi}{2}(1-\gamma)\big)\Big]}{\cos\phi\gamma-5\cos\phi+4\cos\Phi_{AC}^{\sigma}+4i\sin\phi}   
\label{eq:trm}
\end{equation}
\end{widetext}
as a function of energy ($\phi=2\pi k\r$), SOC
($\Phi_{AC}^{\sigma}(\beta)$) and asymmetry \big($\gamma=(1-R)/(1+R)$
\big), where $R$ stands for the ratio of lower and upper ring arm lengths (see
Fig.\ref{fig:tiltangle1}). 

In the following discussion of transmission and conductance, spin index $\sigma$
refers to the spinors in
the ring eigenstates in Eq. \eqref{eq:es}, whereas  the standard
spinor basis (eigenvectors of $\sigma_{z})$ are labeled by $s$. 

\section{Geometry and SOC dependence of  transmission zeros \label{sec:3}}

\subsection{Free system $(\beta=0)$}
\begin{figure}[!h]
\begin{center}
    \footnotesize
  \psfrag{0}[][][0.6]{$0$}
  \psfrag{p1}[][][0.6]{$$}
  \psfrag{p2}[][][0.6]{$2\pi$}
    \psfrag{p3}[][][0.6]{$$}
    \psfrag{p4}[][][0.6]{$4\pi$}
    \psfrag{p5}[][][0.6]{$$}
    \psfrag{p6}[][][0.6]{$6\pi$}
    \psfrag{p7}[][][0.6]{$$}
    \psfrag{p8}[][][0.6]{$8\pi$}
    \psfrag{p9}[][][0.6]{$$}
    \psfrag{p10}[][][0.6]{$10\pi$}
    \psfrag{p11}[][][0.6]{$$}
    \psfrag{k}[t][c]{$\phi$}
    \psfrag{Tu}{$G$}
    \psfrag{ga}[][r][0.8]{$R=1$\qquad}
    \psfrag{gb}[][r][0.8]{$R=1/2$\qquad}
    \psfrag{gc}[][r][0.8]{$R\approx0.55$\qquad}
        \includegraphics[width=3.2in]{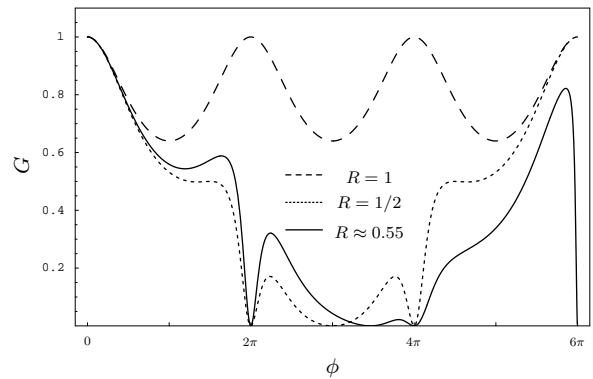}
        \end{center}
    \caption{Conductance for $\beta=0$ in symmetric ($R=1$) and
  asymmetric system ($R=1/2$ and $R=(2\sqrt{3}-1)/(2\sqrt{3}+1)\approx0.55$. For the lead-ring coupling assumed in the
  present model ($\epsilon=4/9$), transmission is not perfect even in
  case of equal branch length. In the asymmetric system, periodical
  transmission zeros appear.} 
    \label{fig:nosoc}
\end{figure}

The transmission function in Eq.\eqref{eq:trm} displays a peculiar
resonant behavior characterized by a set of zeros and poles. The
transmission zeros are obtained from Eq.\eqref{eq:trm} as the
solution of 
\begin{equation}
\sin\Big(\frac{\phi}{2}(\gamma-1)\Big)=e^{-i\Phi_{AC}^{\sigma}}\sin\Big(\frac{\phi}{2}(\gamma+1)\Big).
\label{eq:z}\end{equation}
For $\beta=0$, the phase factor equals unity, and Eq. \eqref{eq:z} simplifies to 
\begin{equation}
\cos\Big(\frac{\phi}{2}\gamma\Big)\sin\Big(\frac{\phi}{2}\Big)=0,\end{equation}
what yields zeros at
\begin{equation}
\phi_{0,1}=2m\pi \quad \textrm{and}\quad \phi_{0,2}=(2m+1)\pi/\gamma,\quad m\in\mathbbm{Z}.
\end{equation}
Obviously, there are two types of zeros. The zeros of the first kind
at $\phi_{0,1}$ correspond to the eigenstates of the closed ring,
whereas the zeros of second type at $\phi_{0,2}$ are given by the
geometry dependent interference condition for nodes at the right
junction\cite{yi:03} and appear only in an asymmetric configuration
($\gamma\neq0$). 
The poles related to transmission resonances are determined by 
\begin{equation}
\cos\phi\gamma-5\cos\phi+4\cos\Phi_{AC}^{\sigma}+4i\sin\phi=0.
\end{equation}
Fig. \ref{fig:nosoc} shows the conductance in absence of SOC
($\beta=0$) for symmetry ($R=1$) and asymmetry parameters $R=1/2$ and $R=(2\sqrt{3}-1)/(2\sqrt{3}+1)\approx0.55$. The
oscillation in the conductance for the symmetric configuration is due
to the coupling of lead and ring, which does not correspond to perfect
transmission and therefore leads to resonances as a consequence of
backscattering effects\cite{buettiker:84}. These resonances however do
not give rise to conductance zeros: from Eq.\eqref{eq:trm} follows
that zeros and poles of the conductance compensate each other and yield a finite
value. In the asymmetric ring ($R=1/2$, $R\approx0.55$), both types of zeros appear. 

\begin{figure}[htbp]
\pspicture(0,0)(5,6.0)
        \footnotesize
          \psfrag{p1}[][][0.6]{$$}
    \psfrag{p2}[][][0.6]{$$}
    \psfrag{p3}[][][0.6]{$$}
    \psfrag{p4}[][][0.6]{$$}
    \psfrag{p5}[][][0.6]{$$}
    \psfrag{p6}[][][0.6]{$$}
    \psfrag{0}[][][0.6]{$$}
     \psfrag{t}[][][0.6]{$-\pi$}
    \psfrag{u}[][][0.6]{$0$}
    \psfrag{v}[][][0.6]{$\pi$}
    \psfrag{re}[][]{$$}
    \psfrag{im}[][]{$Im~\phi$}
         \rput(2.2,4.6){\includegraphics[width=3.5in]{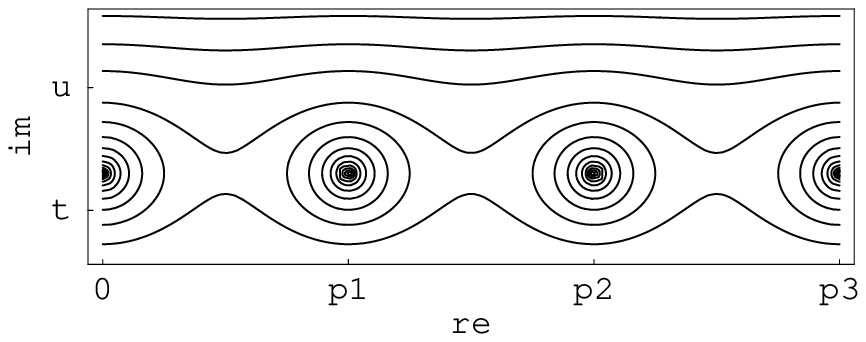}}
      \psfrag{p2}[][][0.6]{$$}
    \psfrag{p3}[][][0.6]{$$}
    \psfrag{p4}[][][0.6]{$$}
    \psfrag{p5}[][][0.6]{$$}
    \psfrag{p6}[][][0.6]{$$}
    \psfrag{0}[][][0.6]{$$}
     \psfrag{t}[][][0.6]{$-\pi$}
    \psfrag{u}[][][0.6]{$0$}
    \psfrag{v}[][][0.6]{$\pi$}
    \psfrag{re}[][]{$$}
    \psfrag{im}[][]{$Im~\phi$}
                \rput(2.2,2.3){\includegraphics[width=3.5in]{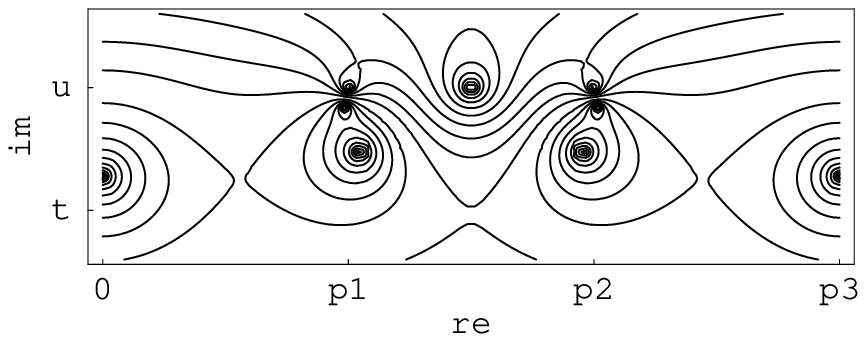}}
    \psfrag{p1}[][][0.6]{$2\pi$}
    \psfrag{p2}[][][0.6]{$4\pi$}
    \psfrag{p3}[][][0.6]{$6\pi$}
    \psfrag{p4}[][][0.6]{$8\pi$}
    \psfrag{p5}[][][0.6]{$10\pi$}
    \psfrag{p6}[][][0.6]{$12\pi$}
     \psfrag{re}[][]{$Re~\phi$}
     \psfrag{im}[][]{$Im~\phi$}
     \psfrag{0}[][][0.6]{$0$}
      \rput(2.2,.0){\includegraphics[width=3.5in]{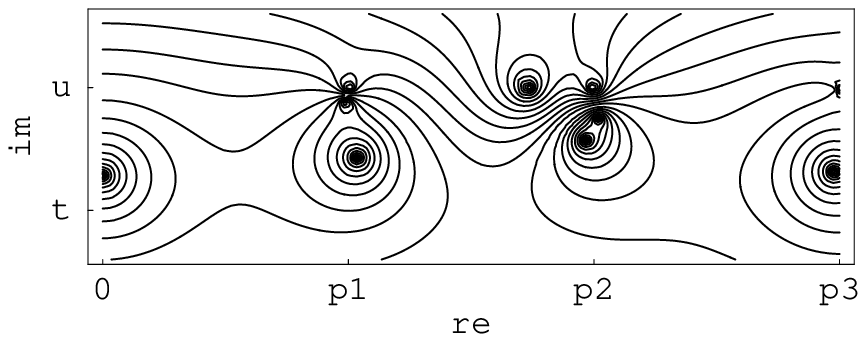}}
    \uput[0](-1.8,5.7){$(a)$}
    \uput[0](-1.8,3.4){$(b)$}
    \uput[0](-1.8,1.2){$(c)$}
    \endpspicture
        \vspace{1.2cm}
        \caption{Zero-pole structure of $G$ in the complex plane for
        $\beta=0$ and (a) $R=1$, (b) $R=1/2$, (c) $R\approx0.55$. The zeros lie on the real axis, whereas the poles have
	a finite imaginary part.} 
        \label{fig:polesnosoc}
\end{figure}

\begin{figure}[htbp]
\pspicture(0,0)(5,3.6)
        \footnotesize
          \psfrag{q1}[][][0.6]{$$}
    \psfrag{q2}[][][0.6]{$$}
    \psfrag{q3}[][][0.6]{$$}
    \psfrag{q4}[][][0.6]{$$}
    \psfrag{q5}[][][0.6]{$$}
    \psfrag{q6}[][][0.6]{$$}
    \psfrag{0}[][][0.6]{$$}
     \psfrag{t}[][][0.6]{$-\pi$}
    \psfrag{u}[][][0.6]{$0$}
    \psfrag{v}[][][0.6]{$\pi$}
    \psfrag{a}[b]{$(a)$}
    \psfrag{b}[b]{$(b)$}
     \psfrag{rf}[c][r]{$$}
    \psfrag{im}[c][c]{$Im~\phi$}
    \rput(2.2,2.45){\includegraphics[width=3.5in]{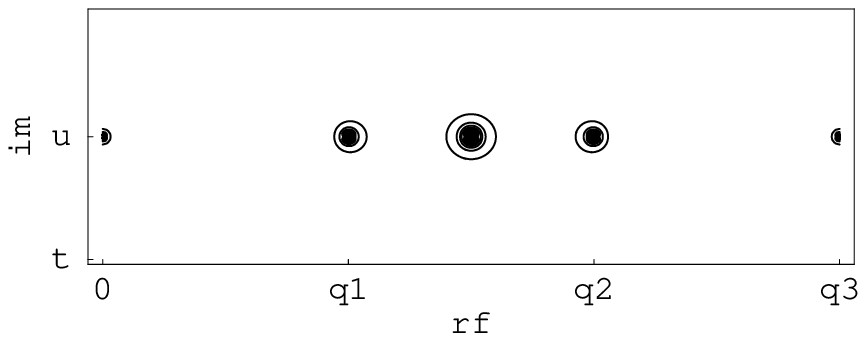}}
                 \psfrag{p1}[][][0.6]{$2\pi$}
    \psfrag{p2}[][][0.6]{$4\pi$}
    \psfrag{p3}[][][0.6]{$6\pi$}
    \psfrag{p4}[][][0.6]{$8\pi$}
    \psfrag{p5}[][][0.6]{$10\pi$}
    \psfrag{p6}[][][0.6]{$12\pi$}
     \psfrag{re}[c][r]{$Re~\phi$}
     \psfrag{im}[c][c]{$Im~\phi$}
     \psfrag{0}[][][0.6]{$0$}
     \rput(2.15,.0){\includegraphics[width=3.8in]{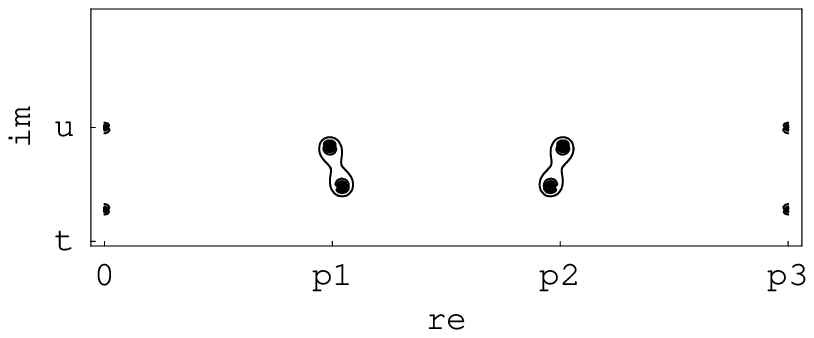}}
                \uput[0](-1.8,3.6){$(a)$}
                \uput[0](-1.8,1.2){$(b)$}
                \endpspicture
        \vspace{1.2cm}
        \caption{(a) zeros and (b) poles of $G$ in the asymmetric case ($R=1/2$)
	for $\beta=0$.}
        \label{fig:zeropole}
\end{figure}

By examination of the transmission amplitude in the complex energy
plane we find a certain connection between the conductance zeros and transmission
resonances. Zeros on the real axis are accompanied by nearby poles in complex plane
(Fig. \ref{fig:polesnosoc} ) and
\ref{fig:zeropole}).  Fig. \ref{fig:zeropole} shows zeros (a) and poles
(b) at $R=1/2$ separately .
A similar feature is known from the quantum
waveguide systems with an attached resonator \cite{porod:94}. In the
present case a pair of poles is associated with each zero of the
first kind at $\phi_{0,1}$. The real part of the energies of the zeros and poles are not exactly
identical, which results in an asymmetric shape of the resonance (Fano type) \cite{fano:61}. 
These characteristic features can be clearly observed in Fig. \ref{fig:nosoc}, e.g. at
$\phi_{0,1}=2\pi$.  Note that at $\phi=0$ and $\phi=6\pi$  both
numerator and denominator of the transmission amplitude vanish simultaneously for $R=1/2$,
such that they annihilate at these places, as can be easily observed in
Fig. \ref{fig:nosoc}. 
\subsection{Finite Rashba-SOC $(\beta\neq0)$}
\begin{figure}[htbp]
\pspicture(0,0)(5,13)
  \footnotesize
  \psfrag{p1}[][][0.6]{$$}
  \psfrag{p2}[][][0.6]{$2\pi$}
    \psfrag{p3}[][][0.6]{$$}
    \psfrag{p4}[][][0.6]{$4\pi$}
    \psfrag{p5}[][][0.6]{$$}
    \psfrag{p6}[][][0.6]{$6\pi$}
    \psfrag{p7}[][][0.6]{$$}
    \psfrag{p8}[][][0.6]{$8\pi$}
    \psfrag{p9}[][][0.6]{$$}
    \psfrag{p10}[][][0.6]{$10\pi$}
    \psfrag{p11}[][][0.6]{$$}
    \psfrag{a}[b]{$(a)$}
    \psfrag{b}[b]{$(b)$}
    \psfrag{k}[t][c]{$\phi$}
    \psfrag{Tu}{$|T_{\uparrow s}|^{2}$}
    \psfrag{ga}[][r][0.8]{$s=\uparrow$}
    \psfrag{gb}[][r][0.8]{$s=\downarrow$}
    \psfrag{q}[][][0.6]{$$}
    \psfrag{l}[][][0.6]{$$}
            \rput(2.2,11.0){\includegraphics[width=3.2in]{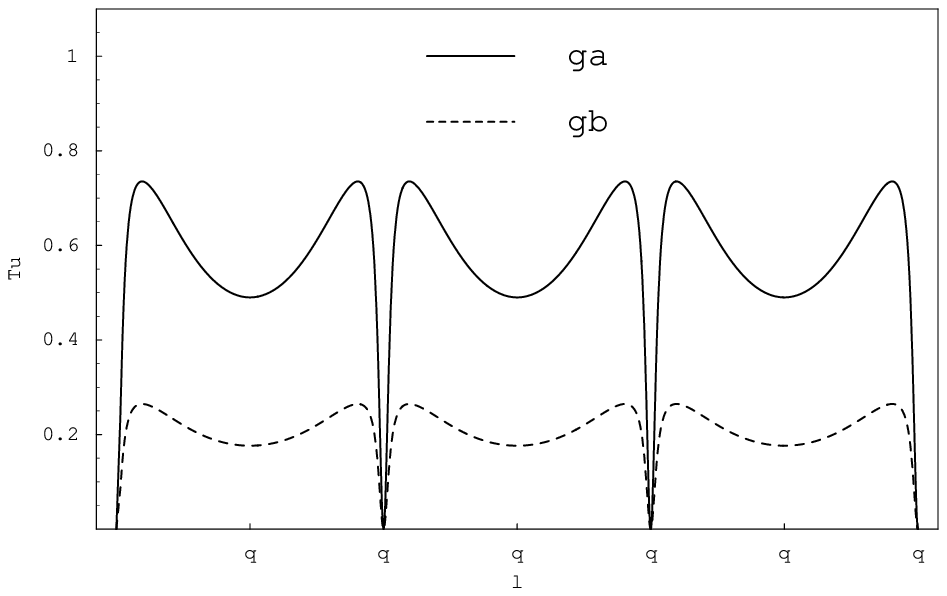}}
            \rput(2.2,6.75){\includegraphics[width=3.2in]{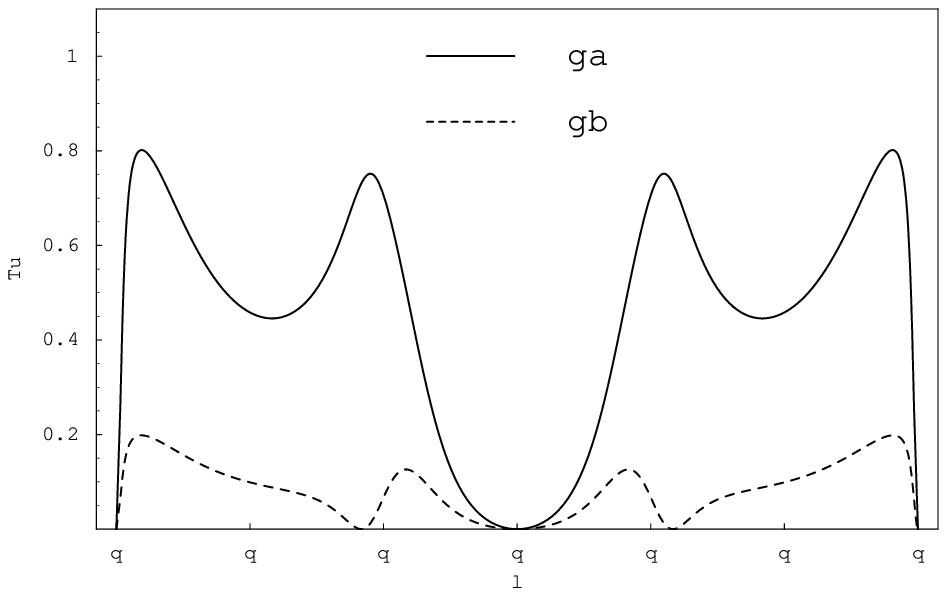}}
            \rput(2.2,2.50){\includegraphics[width=3.2in]{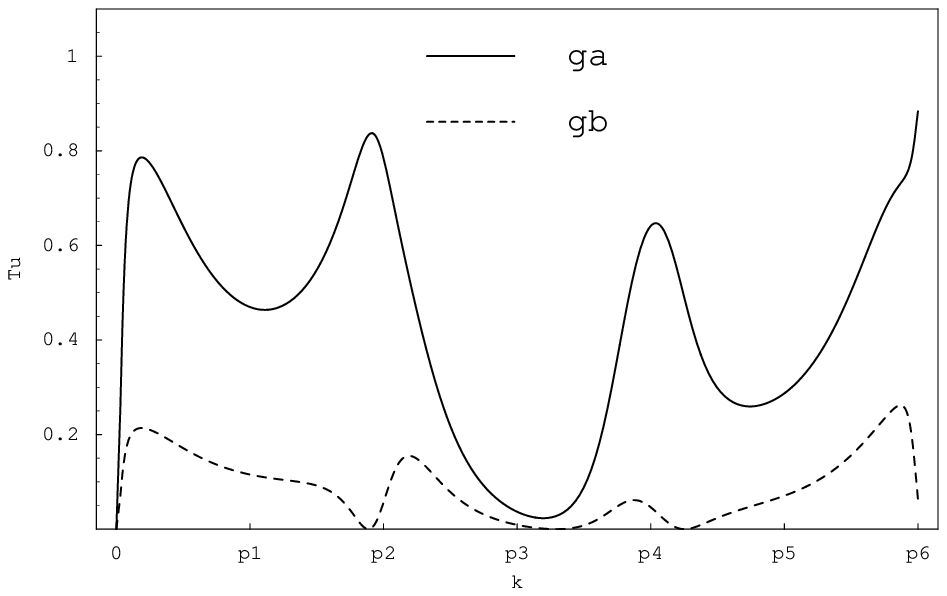}}
            \uput[0](-2.0,13.){$(a)$}
            \uput[0](-2.0,8.8){$(b)$}
            \uput[0](-2.0,4.5){$(c)$}
   \endpspicture
 \caption{Transmission probability for nonzero SOC $\beta=0.6$ from $s=\uparrow$
 into $s=\uparrow$ and $s=\downarrow$
  spin channels for symmetric  ((a) $R=1$) and  asymmetric system
  ((b) $R=1/2$, (c) $R\approx0.55$). In the symmetric AC-ring, spin-orbit interaction causes the
  appearance of transmission zeros, in the asymmetric configuration
  however, the latter are partially lifted. For particular asymmetry
  ratios $R$, the geometry dependent zeros persist.} 
    \label{fig:soc}
\end{figure}

\begin{figure}[htbp]
\pspicture(0,0)(5,6.0)
        \footnotesize
          \psfrag{p1}[][][0.6]{$$}
    \psfrag{p2}[][][0.6]{$$}
    \psfrag{p3}[][][0.6]{$$}
    \psfrag{p4}[][][0.6]{$$}
    \psfrag{p5}[][][0.6]{$$}
    \psfrag{p6}[][][0.6]{$$}
    \psfrag{0}[][][0.6]{$$}
     \psfrag{t}[][][0.6]{$-\pi$}
    \psfrag{u}[][][0.6]{$0$}
    \psfrag{v}[][][0.6]{$\pi$}
    \psfrag{re}[][]{$$}
    \psfrag{im}[][]{$Im~\phi$}
         \rput(2.2,4.6){\includegraphics[width=3.5in]{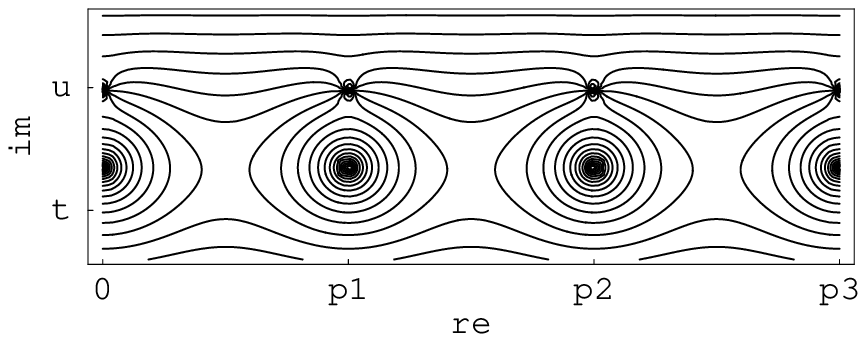}}
      \psfrag{p2}[][][0.6]{$$}
    \psfrag{p3}[][][0.6]{$$}
    \psfrag{p4}[][][0.6]{$$}
    \psfrag{p5}[][][0.6]{$$}
    \psfrag{p6}[][][0.6]{$$}
    \psfrag{0}[][][0.6]{$$}
     \psfrag{t}[][][0.6]{$-\pi$}
    \psfrag{u}[][][0.6]{$0$}
    \psfrag{v}[][][0.6]{$\pi$}
    \psfrag{re}[][]{$$}
    \psfrag{im}[][]{$Im~\phi$}
         \rput(2.2,2.3){\includegraphics[width=3.5in]{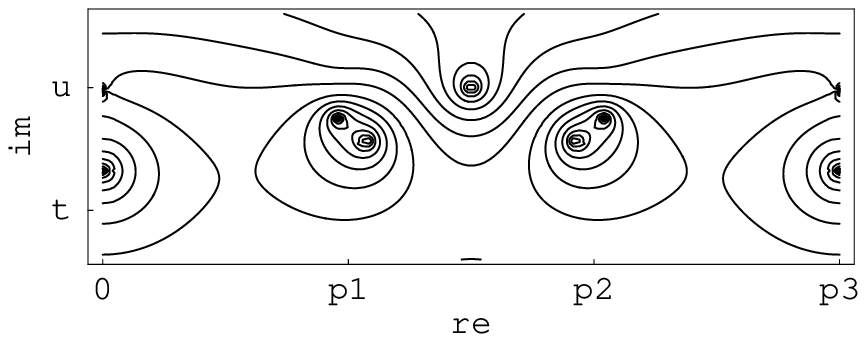}}
    \psfrag{p1}[][][0.6]{$2\pi$}
    \psfrag{p2}[][][0.6]{$4\pi$}
    \psfrag{p3}[][][0.6]{$6\pi$}
    \psfrag{p4}[][][0.6]{$8\pi$}
    \psfrag{p5}[][][0.6]{$10\pi$}
    \psfrag{p6}[][][0.6]{$12\pi$}
     \psfrag{re}[][]{$Re~\phi$}
     \psfrag{im}[][]{$Im~\phi$}
     \psfrag{0}[][][0.6]{$0$}
                \rput(2.2,.0){\includegraphics[width=3.5in]{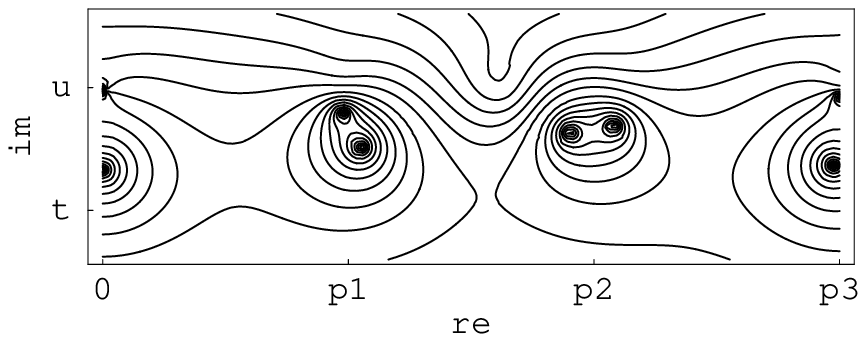}}
                \uput[0](-1.8,5.7){$(a)$}
                \uput[0](-1.8,3.4){$(b)$}
                \uput[0](-1.8,1.2){$(c)$}
                \endpspicture
        \vspace{1.2cm}
        \caption{Zero-pole structure of $G$ in the complex plane for
        $\beta=0.6$ and (a) $R=1$, (b) $R=1/2$, (c) $R\approx0.55$.} 
        \label{fig:polessoc}
\end{figure}

\begin{figure}[htbp]
\pspicture(0,0)(5,3.7)
        \footnotesize
          \psfrag{p1}[][][0.6]{$$}
    \psfrag{p2}[][][0.6]{$$}
    \psfrag{p3}[][][0.6]{$$}
    \psfrag{p4}[][][0.6]{$$}
    \psfrag{p5}[][][0.6]{$$}
    \psfrag{p6}[][][0.6]{$$}
    \psfrag{0}[][][0.6]{$$}
     \psfrag{t1}[][][0.6]{$-\pi$}
    \psfrag{u1}[][][0.6]{$0$}
    \psfrag{v1}[][][0.6]{$\pi$}
    \psfrag{a}[b]{$(a)$}
    \psfrag{b}[b]{$(b)$}
     \psfrag{re}[c][r]{$$}
    \psfrag{im1}[c][c]{$Im~\phi$}
        \rput(2.2,2.3){\includegraphics[width=3.6in]{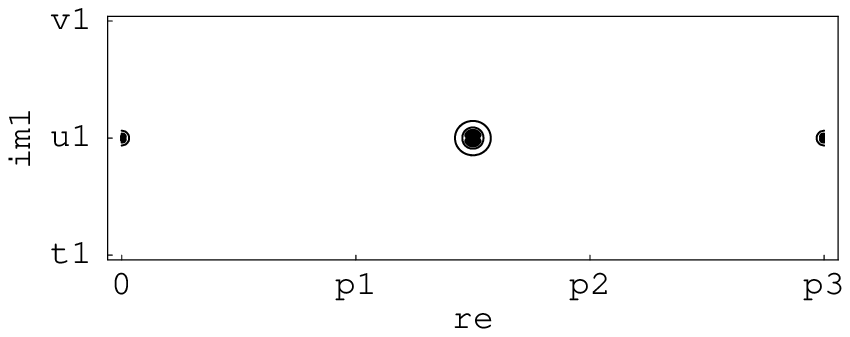}}
                 \psfrag{p1}[][][0.6]{$2\pi$}
    \psfrag{p2}[][][0.6]{$4\pi$}
    \psfrag{p3}[][][0.6]{$6\pi$}
    \psfrag{p4}[][][0.6]{$8\pi$}
    \psfrag{p5}[][][0.6]{$10\pi$}
    \psfrag{p6}[][][0.6]{$12\pi$}
     \psfrag{re}[c][r]{$Re~\phi$}
     \psfrag{im1}[c][c]{$Im~\phi$}
     \psfrag{0}[][][0.6]{$0$}
     \rput(2.2,.0){\includegraphics[width=3.6in]{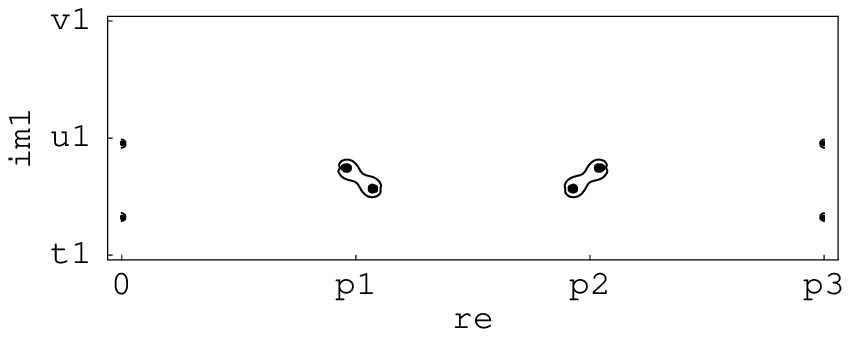}} 
                \uput[0](-1.8,3.4){$(a)$}
                \uput[0](-1.8,1.2){$(b)$}
                \endpspicture
        \vspace{1.2cm}        \caption{(a) zeros and (b) poles of $G$ in the
	asymmetric case ($R=1/2$) for $\beta=0.6$.}
        \label{fig:zeropolesoc}
\end{figure}

\begin{figure}[htbp]
\pspicture(0,0)(2,2.2)
\footnotesize
 \psfrag{q1}[][][0.6]{$$}
    \psfrag{q2}[][][0.6]{$$}
    \psfrag{q3}[][][0.6]{$$}
    \psfrag{q4}[][][0.6]{$$}
    \psfrag{q5}[][][0.6]{$$}
    \psfrag{q6}[][][0.6]{$$}
    \psfrag{0}[][][0.6]{$0$}
     \psfrag{t}[][][0.6]{$-\pi$}
    \psfrag{u}[][][0.6]{$0$}
    \psfrag{v}[][][0.6]{$\pi$}
     \psfrag{t1}[][][0.6]{$$}
    \psfrag{u1}[][][0.6]{$$}
    \psfrag{v1}[][][0.6]{$$}
    \psfrag{re1}[c][c]{$$}
    \psfrag{im1}[c][r]{$$}
     \psfrag{a}[][]{$a)$}
    \psfrag{b}[][]{$b)$}
     \psfrag{c}[][]{$c)$}
    \psfrag{d}[][]{$d)$}
    \psfrag{p1}[][][0.6]{$2\pi$}
    \psfrag{p2}[][][0.6]{$4\pi$}
    \psfrag{p3}[][][0.6]{$6\pi$}
    \psfrag{p4}[][][0.6]{$8\pi$}
    \psfrag{p5}[][][0.6]{$10\pi$}
    \psfrag{p6}[][][0.6]{$12\pi$}
     \psfrag{re}[c][c]{$Re~\phi$}
     \psfrag{im}[c][c]{$Im~\phi$}
\includegraphics[bb=140 15 280 60,width=2.3in]{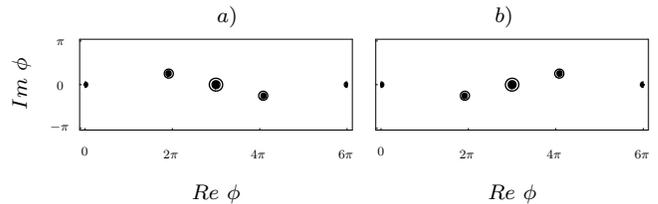}
\uput[0](-3.7,2.2){$a)$}
\uput[0](0,2.2){$b)$}
\endpspicture
\caption{Zeros of a) $T_{\uparrow}$ and b) $T_{\downarrow}$ for
  $\beta=0.6$ and $R=1/2$. SOC shifts the zeros away from the real
  axis, the direction of the shift depending on the spin.} 
\label{fig:zeroshift}
\end{figure}

There are two remarkable features in the transmission characteristics arising as effects of SOC.
The first is the finite transmission probability in the spin channel
opposite to the incident spin orientation. This is the result of spin
precession along the ring branches due to SOC as considered in
Ref. \onlinecite{bulgakov:02}. The conductance zeros in the opposite
channel correspond to a frequency of precession which reproduces the
incident spin orientation at the right junction. The second aspect, and
the one on which we will concentrate in the following, is the lifting
of certain conductance zeros in the incident channel. These features
can be observed in Fig. \ref{fig:soc} where the transmission for
finite SOC is displayed.  
It is instructive to analyze the modification of the transmission
amplitude in the complex energy plane. Fig.\ref{fig:polessoc} shows
the lifting of the zeros of the first kind as well as the emergence of
zeros that were canceled by poles in the free system. The shifting of
zeros and poles away from the real axis is displayed in
Fig. \ref{fig:zeropolesoc}. In the conductance, the zeros of the first
kind appear no longer. They are still present in the up- and down
transmission amplitudes, but different spin components are shifted in
opposite directions, as it is shown in Fig. \ref{fig:zeroshift}. 
\newline
\newline
To study the behavior of the transmission zeros under the influence of
SOC, an expansion around the zeros in the AC-phase
$\Phi_{AC}^{\sigma}~(mod~2\pi)$ of the transmission probability
$\mathscr{T}_{\sigma}=|T_{\sigma}|^{2}$ is performed:   
\begin{align}
\mathscr{T}_{\sigma}(\Phi_{AC}^{\sigma})&
=8\csc(\pi m\gamma)\left(\Phi_{AC}^{\sigma}\right)^{2}+
O\big[\left(\Phi_{AC}^{\sigma}\right)^{3}\big]\nonumber \\
&\quad\mathrm{at}\quad\phi_{0,1}=2m\pi,\label{eq:expan1}\\
\nonumber\\
\mathscr{T}_{\sigma}(\Phi_{AC}^{\sigma})&
=\frac{16\left[1+\cos\big((2m+1)\pi/\gamma\big)\right]}{\left[5-3
\cos\big((2m+1)\pi/\gamma\big)\right]^{2}}\left(\Phi_{AC}^{\sigma}\right)^{2}\nonumber\\ 
&+O\big[\left(\Phi_{AC}^{\sigma}\right)^{3}\big]\nonumber\\
&\quad\mathrm{at}\quad\phi_{0,2}=(2m+1)\pi/\gamma.\label{eq:expan2}
\end{align}
Eq.\eqref{eq:expan1} shows that the zeros of the first type are
removed by the action of SOC for all values of $\gamma$. For the zeros
of the second type however there are geometries where the zeros
persist even in presence of the interaction. From
Eq.\eqref{eq:expan2} follows the geometry condition for persistent
zeros: 
\begin{align}
\gamma_{per}=\frac{2m+1}{2n+1}\quad\Leftrightarrow\quad R_{per}=\frac{m+n+1}{n-m},\\
\nonumber\\
n,m\in\mathbb{Z},~n>m.\nonumber
\end{align}
\newline

\section{Analogy to AB-ring and symmetry argument\label{sec:4}}
\begin{figure}[htbp]
   \begin{flushleft}
    \psfrag{a}{$\Phi=0$}
    \psfrag{b}{$\Phi\neq0$}
    \psfrag{E}{E}
    \psfrag{g}{$=$}
    \psfrag{ng}{$\neq$}
    \psfrag{r}{resonant eigenstates}
    \psfrag{t}{TRS broken}
        \includegraphics[width=2.0in]{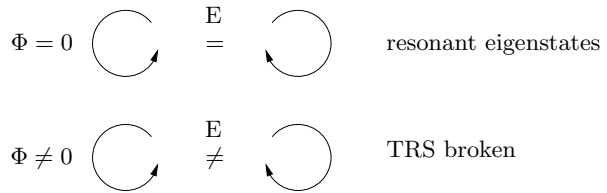}
   \end{flushleft}
    \caption{Resonant states and broken symmetry for AB-ring (E=energy).}
    \label{fig:symmetry}
\end{figure}

It was shown \cite{wakabayashi:01,wakabayashi:00b} for the AB-ring that zero conductance
energies belong to states of vanishing vorticity, i.e. the circular
currents in the loop system change sign at these energies. The zero
conductance resonances can therefore be regarded as the signatures of
destructive 
interference resulting from the superposition of circular currents
of opposite chirality corresponding to degenerate resonant states
of the loop system. The possibility of superposition is due to the
degeneracy of the two chiral states as  a consequence of time reversal
symmetry in absence of external fields. A magnetic field,
respectively the resulting flux $\Phi$ through the
loop, destroys this degeneracy as a consequence of 
broken TRS (Fig. \ref{fig:symmetry}).

\begin{figure}[htbp]
    \begin{flushleft}
    \psfrag{a}{$\beta=0$}
    \psfrag{b}{$\beta\neq0$}
    \psfrag{E}{E}
    \psfrag{g}{$=$}
    \psfrag{ng}{$\neq$}
    \psfrag{r}{resonant eigenstates}
    \psfrag{t}{TRS preserved}
    \psfrag{p}{parity broken}
    \psfrag{u}{$\uparrow$}
    \psfrag{d}{$\downarrow$}
    \psfrag{ud}{$\uparrow,\downarrow$}
        \includegraphics[width=2.0in]{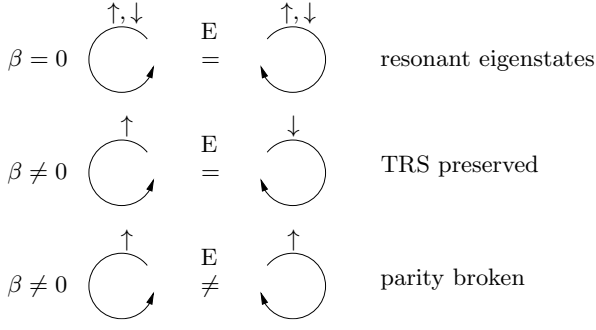}
    \end{flushleft}
    \caption{Resonant states and broken parity for ring subject to Rashba SOC. }
    \label{fig:symmetrysoc}
\end{figure}

In the present case of a one-dimensional ring subject to Rashba SOC, the
role of the magnetic flux $\Phi$ is played by the Rashba term depending
on the coupling $\beta$. In fact, the transmission function in
Eq.\eqref{eq:trm} equals the expression obtained in Ref.
\onlinecite{wakabayashi:01} for the asymmetric
AB-ring, except that the AB-phase $\Phi_{AB}=2\pi\Phi/\Phi_{0}$ is
replaced by the (spin dependent) AC-phase $\Phi_{AC}^{\sigma}$. In
analogy to the AB-ring, 
there are conductance zeros due to resonant states of different
chirality for the free system at $\beta=0$. At finite SOC,
configurations of opposite spin \emph{and} chirality are still degenerate as a
consequence of time reversal symmetry: with the time reversal operator
given by $\hat{T}=-i\sigma_{y} \hat K$, where $\hat K$ is the operator
for complex conjugation, and using the relations in Eq.\eqref{eq:qn},
we find  
\begin{equation}
\hat T \Psi_{n,\lambda}^{\sigma}=-\sigma\Psi_{n,-\lambda}^{-\sigma}.
\end{equation}
For a fixed spin orientation however, states of opposite chirality are no longer
degenerate as parity is broken for $\beta\neq0$. The situation
with SOC is illustrated in Fig. \ref{fig:symmetrysoc}.
The vanishing of circular currents corresponding to time reversed
degenerate states is easily derived: from Eqs. \eqref{eq:v0},
\eqref{eq:vso} follow the currents
\begin{align}
j_{\lambda}^{\sigma}&
=2\left(n_{\lambda}^{\sigma}+\sin^{2}\Big(\frac{\theta}{2}+\frac{\pi}{4}(1-\sigma)\Big)\right)+\sigma\beta\sin\theta\\
&\sigma=\pm1,\quad\lambda=\pm. \nonumber
\end{align}
The total circular current of time reversed states has to vanish such that
\begin{equation}
j_{tot}\left(\Psi_{\lambda}^{\sigma},\Psi_{-\lambda}^{-\sigma}\right)
=2\left(n_{\lambda}^{\sigma}+1+n_{-\lambda}^{-\sigma}\right)\equiv0\quad\forall\beta,
\end{equation}
whereas the total circular current for states of a equal spin
disappear only for $\beta \to 0$,
\begin{align}
j_{tot}\left(\Psi_{+}^{\uparrow},\Psi_{-}^{\uparrow}\right)&
=2\left(n_{+}^{\uparrow}+n_{-}^{\uparrow}+2\sin^{2}\frac{\theta}{2}+\beta\sin\theta\right),\\
j_{tot}\left(\Psi_{+}^{\downarrow},\Psi_{-}^{\downarrow}\right)&
=2\left(n_{+}^{\downarrow}+n_{-}^{\downarrow}+2\cos^{2}\frac{\theta}{2}-\beta\sin\theta\right).    
\end{align}
The symmetry breaking analogy between AB-rings and rings subject to
Rashba-SOC appears already in the corresponding Hamiltonians and their
symmetries. For Rashba SOC, the normalized magnetic flux
$\Phi/\Phi_{0}$ breaking time reversal symmetry in the AB-ring  is
replaced by the spin dependent vector potential $A(\varphi)$ which
respects the TRS of $\hat{H}$, i.e. 
\begin{equation}
\big[\hat H,\hat
T\big]_{\Psi_{n,\lambda}^{\sigma}}=\left(n+1+
\Phi_{AC}^{-\sigma}\right)^{2}-\left(n-\Phi_{AC}^{\sigma}\right)^{2}\equiv0\quad\forall 
\beta, 
\end{equation} 
but changes under spin reversal, and which is related to the Aharonov-Casher
phase by Eq.\eqref{eq:eigenenergy} for the eigenenergies. The main 
results of this analysis are summarized in Tab. \ref{tab:analogy}. 
\begin{table}[h!] 
    \begin{flushleft}
        \begin{tabular}[c]{l|l|l}      
        &\hspace{0.3in}AB-ring&\hspace{0.1in} ring with Rashba-SOC\\
        \hline    
        $\begin{array}{l}\textrm{ext.}\\ \textrm{field}\\
        \end{array}$&
$\vec{B}=(0,0,B_{z})$&$\vec{E}=(0,0,E_{z})$\\
        \hline
        $\begin{array}{l}\textrm{Hamil-}\\\textrm{tonian}\\\end{array}$&
$\hat{H}=\frac{1}{2m\r^{2}}\left(\frac{\hbar}{i}\frac{\partial}{\partial
        \varphi}+\frac{\Phi}{\Phi_{0}}\right)^{2}$&$\hat{H}=\frac{1}{2m\r^{2}}
\left(\frac{\hbar}{i}\pf{\varphi}+A(\varphi)\right)^{2}$\\  
        &$\Phi_{0}=\frac{hc}{e}$&$A(\varphi)=\frac{\beta\hbar^{2}}{2}\sigma_{r}(\varphi)$\\
        &&\\
        \hline
        $\begin{array}{l} 
        \textrm{broken} \\ 
        \textrm{symm.}\\ 
        \end{array}$& time reversal \quad $\hat{T}$&spin parity \quad $\hat{P}_{s}$\\
        &$\begin{array}{lcl}[\hat{H},\hat{T}]&=
&\frac{2\hbar k}{m}\frac{\Phi}{\Phi_{0}}\\&=&0\Leftrightarrow\Phi=0
        \end{array}$&$\begin{array}{lcl}[\hat{H},\hat{P}_{s}]&=&-i\beta 
\sin\varphi\sigma_{z}\\&=&0\Leftrightarrow\beta=0\end{array}$\\ 
        \end{tabular}
    \end{flushleft}
    \caption{Symmetry breaking analogy between AB- and Rashba-rings.}
    \label{tab:analogy}
\end{table}

In Ref. \onlinecite{kim:02}, a relation was established between the
breaking of TRS by a magnetic field in an AB-ring and the location of
the transmission zeros in the complex plane. It was shown that real
zeros appear if the flux is an integer or half integer multiple of the
flux quantum $\Phi_{0}$, and are shifted off the real axis for other
flux values. Due to the analogy to the AB-ring, the behavior of the
transmission zeros of the ring subject to Rashba-SOC follows the same
rules, now depending on the value of the AC-phase. This implies the
periodical dependence of transmission properties on the value of the
SOC-constant $\beta$: real transmission zeros demand a (half) integer
AC-Phase, $\Phi_{AC}^{\sigma}/2\pi=(2m+1)/2$, $m\in\mathbbm{Z}$, which is
satisfied by\cite{frustaglia:04} $\beta=\sqrt{4(m+1)^{2}-1}$. 

It was shown by Lee and co-workers that conductance zeros occur
generically in quasi-1D systems if time reversal is a
symmetry\cite{lee:99,lee:01}. In the proof, they used the constraints
laid upon the elements of the scattering matrix describing the system of
spinless particles by the symmetry and unitarity requirement. In the case of
particles with spin, these conditions are reproduced only in the 
presence of time reversal symmetry {\it and} parity, apart from
special situations (geometries). 

As in Ref. \onlinecite{molnar:04}, it is possible to combine the AB- and
AC-effects by the inclusion of a finite magnetic flux in the Hamiltonian
\eqref{eq:Hamiltonian},
\begin{equation}
H=\left(-i\pf{\varphi}+\frac{\beta}{2}\sigma_{r}-\frac{\Phi}{\Phi_{0}}\right)^{2}.\label{eq:HamiltonianFlux} 
\end{equation}
Eq.\eqref{eq:n} becomes\cite{molnar:04}
\begin{equation}
n_{\lambda}^{\sigma}(k)=\lambda k\r+\big(\Phi_{AC}^{\sigma}+\Phi_{AB}\big)/2\pi.
\label{eq:nflux}
\end{equation} 
Thus, the AB-effect contributes just a spin independent additive phase, i.e. in eq.
\eqref{eq:trm}, the AC-phase has to be replaced by the sum of AB- and AC-phases.
\section{Conclusions\label{sec:5}}
In summary, we have shown that zero conductance resonances
appearing as a signature of interfering resonant states of the loop
system, and as a consequence of its asymmetry, behave in a similar way
under the influence of a magnetic flux through the loop as in
presence of a perpendicular electric field generating Rashba
spin-orbit coupling. Real conductance zeros are lifted by the
influence of these external fields,
being shifted into the complex plane depending on the
value of the AB(AC)-phase. In the case of the magnetic flux, it is the
breaking of time reversal symmetry which destroys the energetic
degeneracy of states with opposite chirality, preventing the
destructive interference leading to the zeros. For Rashba SOC, time
reversal symmetry is respected, but not spin reversal symmetry, 
which again leads
to a chiral dependence in the energy of the loop wave function and
eventually to the lifting of the conductance zeros.

\vskip 0.4 cm

We gratefully acknowledge the financial support of this study 
by a Grant of the Swiss Nationalfonds.  

\vskip 1cm

\appendix

\section{Derivation of the transmission amplitude in the single-electron scattering
picture}

This derivation follows Ref. \onlinecite{molnar:04} and \onlinecite{frustaglia:04}.
System geometry and coordinates are shown in
Fig. \ref{fig:tiltangle1}. Asymmetry is introduced by choosing
different lengths $l_{up}$ and $l_{low}$ for upper and lower branches
of the ring in the figure, and is expressed by means of the asymmetry factor
$R=l_{low}/l_{up}$. This leads to different phases at the left
junction ($A$): 
\begin{equation}
\varphi(A)=\frac{2\pi}{R+1}\equiv\varphi_{A},\qquad \varphi'(A)=\frac{2\pi R}{R+1}\equiv\varphi'_{A}
\end{equation} 
The connection of leads and ring is described by the application of
spin-dependent Griffith's boundary conditions \cite{griffiths:53},
which demand a) continuity of the wave function and b) probability
current conservation  at the junctions (A) and (B)\footnote{In the
  scattering matrix approach of [\onlinecite{buettiker:84}], these conditions
  correspond to a coupling parameter $\epsilon=4/9$, which is below
  the value of $\epsilon=1/2$ for perfect transmission.}. \newline 
To be able to apply the boundary conditions, we need the wave functions of leads and branches. 
The wavefunctions $\Psi_{I}$ and $\Psi_{II}$ for incoming and outgoing
leads respectively, can be expanded in terms of the spinors
$\chi^{\sigma}$ at the junctions, 
\begin{align}
  \Psi_{I}(x)&=\sum_{\sigma=\uparrow,\downarrow}\Psi_{I}^{\sigma}(x)\chi^{\sigma}(\varphi_{A}),\quad x\in[-\infty,0],\\   
\Psi_{II}(x')&=\sum_{\sigma=\uparrow,\downarrow}\Psi_{II}^{\sigma}(x')\chi^{\sigma}(0),\quad
x'\in[0,\infty], 
\end{align}
The expansion coefficients are the orbital wave functions
\begin{align}
        \Psi_{I}^{\sigma}(x)&=i_{\sigma}e^{ikx}+r_{\sigma}e^{-ikx},\\
        \Psi_{II}^{\sigma}(x')&=t_{\sigma}e^{ikx'}
\end{align}
where we assume an incident plane wave from the left with wave number
$k$. The coefficients $i_{\sigma}$ of the incoming wave are chosen
such that $\sum_{\sigma}|i_{\sigma}|^{2}=1$. $r_{\sigma}$ and
$t_{\sigma}$ are the spin dependent reflection and transmission
coefficients, respectively. 
A similar expansion in terms of the ring eigenstates in
(\ref{eq:es}) yields the wave functions $\Psi_{up}$ and
$\Psi_{low}$ of upper and lower branches, respectively: 
\begin{align}
\Psi_{up}(\varphi)&=\sum_{\sigma=\uparrow,\downarrow}\Psi_{up}^{\sigma}(\varphi)\chi^{\sigma}(\varphi),\quad\varphi\in[0,\varphi_{A}],\\
\Psi_{low}(\varphi')&=\sum_{\sigma=\uparrow,\downarrow}\Psi_{low}^{\sigma}(\varphi')\chi^{\sigma}(-\varphi'),\quad\varphi'\in[0,\varphi'_{A}],
\end{align}
with the corresponding orbital components
\begin{align}
        \Psi_{up}^{\sigma}(\varphi)&=\sum_{\lambda=+,-}a_{\lambda}^{\sigma}e^{in_{\lambda}^{\sigma}\varphi},\\
\Psi_{low}^{\sigma}(\varphi')&=\sum_{\lambda=+,-}b_{\lambda}^{\sigma}e^{-in_{\lambda}^{\sigma}\varphi'},           
\end{align}
where $n_{\lambda}^{\sigma}$ is given by Eq.(\ref{eq:n}).

Imposing the boundary conditions mentioned above, it is now possible
to relate the transmission and reflection coefficients $r_{\sigma}$
and $t_{\sigma}$ to the input parameters $i_{\sigma}$. The continuity
conditions for the wave function demand
$\Psi_{II}^{\sigma}(0)=\Psi_{up}^{\sigma}(0)=\Psi_{low}^{\sigma}(0)$
and
$\Psi_{I}^{\sigma}(0)=\Psi_{up}^{\sigma}(\varphi_{A})=\Psi_{low}^{\sigma}(\varphi'_{A})$,
yielding the equations 
\begin{align}
        \sum_{\lambda=+,-}a_{\lambda}^{\sigma}&=\sum_{\lambda=+,-}b_{\lambda}^{\sigma}=t_{\sigma},\label{eq:cont1}\\            \sum_{\lambda=+,-}a_{\lambda}^{\sigma}e^{in_{\lambda}^{\sigma}\varphi_{A}}&=\sum_{\lambda=+,-}b_{\lambda}^{\sigma}e^{-in_{\lambda}^{\sigma}\varphi'_{A}}=r_{\sigma}+i_{\sigma}.
        \label{eq:cont2}
\end{align}
Probability current density conservation requires
$j_{up}^{\sigma}+j_{low}^{\sigma}+j_{I(II)}^{\sigma}=0$ at the junctions. The current densities follow evaluating the expressions derived in Sec.\ref{sec:2} for the wave functions above. The (dimensionless) ring currents read
\begin{align}   j_{up}^{\sigma}(\varphi)=&\frac{1}{2}\Big(\big(\Psi_{up}^{\sigma}\chi^{\sigma}\big)^{\dagger}\big(\hat{v}\Psi_{up}^{\sigma}\chi^{\sigma}\big)\nonumber\\
&+\Psi_{up}^{\sigma}\chi^{\sigma}\big(\hat{v}\Psi_{up}^{\sigma}\chi^{\sigma}\big)^{\dagger}\Big)(\varphi),\\
j_{low}^{\sigma}(\varphi')=&\frac{1}{2}\Big(\big(\Psi_{low}^{\sigma}\chi_{-}^{\sigma}\big)^{\dagger}\big(\hat{v}'\Psi_{low}^{\sigma}\chi_{-}^{\sigma}\big)\nonumber\\
&+\Psi_{low}^{\sigma}\chi_{-}^{\sigma}\big(\hat{v}'\Psi_{low}^{\sigma}\chi_{-}^{\sigma}\big)^{\dagger}\Big)(\varphi'),\end{align}
where $\hat{v}(\varphi)=\hat{v}_{0}(\varphi)+\hat{v}_{SO}(\varphi)$, 
$\hat{v}'(\varphi)=\hat{v}_{0}(\varphi)-\hat{v}_{SO}(\varphi)$ and $\chi_{-}^{\sigma}(\varphi')=\chi^{\sigma}(-\varphi')$. The currents in the leads are given by
\begin{align}   j_{I}^{\sigma}(x)=&\frac{1}{2}\Big(\big(\Psi_{I}^{\sigma}\chi_{A}^{\sigma}\big)^{\dagger}\big(\hat{v}_{0}\Psi_{I}^{\sigma}\chi_{A}^{\sigma}\big)\nonumber\\
&+\Psi_{I}^{\sigma}\chi_{A}^{\sigma}\big(\hat{v}_{0}\Psi_{I}^{\sigma}\chi_{A}^{\sigma}\big)^{\dagger}\Big)(x),\\
j_{II}^{\sigma}(x')=&\frac{1}{2}\Big(\big(\Psi_{II}^{\sigma}\chi_{B}^{\sigma}\big)^{\dagger}\big(\hat{v}_{0}\Psi_{II}^{\sigma}\chi_{B}^{\sigma}\big)\nonumber\\
&+\Psi_{II}^{\sigma}\chi_{B}^{\sigma}\big(\hat{v}_{0}\Psi_{II}^{\sigma}\chi_{B}^{\sigma}\big)^{\dagger}\Big)(x'), 
\end{align}
where $\chi_{A(B)}^{\sigma}=\chi_{\sigma}(\varphi(A(B))$. Using the
equality of the wave function at the junctions and noting that
$\hat{v}_{SO}(\varphi)\chi^{\sigma}(\varphi)=-\hat{v}_{SO}(\varphi)\chi_{-}^{\sigma}(\varphi')$, 
the probability current density conservation condition simplifies to
\begin{align}
\hat{v}_{0}\Psi_{up}^{\sigma}\big|_{\varphi=0\left(\varphi_{A}\right)}&+\hat{v}_{0}\Psi_{low}^{\sigma}\big|_{\varphi'=0\left(\varphi'_{A}\right)}  
+\hat{v}_{0}\Psi_{I(II)}^{\sigma}\big|_{x(x')=0}=0.\nonumber\\
\end{align}
From that follows an additional pair of equations for the coefficients:
\begin{align}
  &\sum_{\lambda=+,-}a_{\lambda}^{\sigma}\frac{n_{\lambda}^{\sigma}}{k\r}-\sum_{\lambda=+,-}b_{\lambda}^{\sigma}\frac{n_{\lambda}^{\sigma}}{k\r}+t_{\sigma}=0,\\ 
&\sum_{\lambda=+,-}a_{\lambda}^{\sigma}e^{in_{\lambda}^{\sigma}\varphi_{A}}\frac{n_{\lambda}^{\sigma}}{k\r}-\sum_{\lambda=+,-}b_{\lambda}^{\sigma}e^{-in_{\lambda}^{\sigma}\varphi'_{A}}\frac{n_{\lambda}^{\sigma}}{k\r}+i_{\sigma}-r_{\sigma}=0.  
\end{align}
Together with Eqs. \eqref{eq:cont1} and \eqref{eq:cont2}, we now have
enough equations to determine the coefficient set $\{r_{\sigma},
~t_{\sigma},~a_{\lambda}^{\sigma},~b_{\lambda}^{\sigma}\}$,
$\lambda=\pm$, for both spin polarizations
$\sigma=\uparrow,\downarrow$ as a function of the input coefficients
$i_{\sigma}$, the incident wave number $k$, ring radius $\r$ and
SOC-constant $\beta$. 
For an incident current from the right, an analogous calculation is
performed with $\{i_{\sigma},~r_{\sigma}\}$ (left lead) and
$\{t_{\sigma},~0\}$ (right lead) replaced by $\{0,~t_{\sigma}'\}$ and
$\{r_{\sigma}',~i_{\sigma}'\}$, respectively. 
This enables us to formulate the scattering matrix of the ring system:
$\vec o=\underline{S}\vec i$, where $\vec o$ stands for outgoing and
$\vec i$ for incoming wave coefficients. The relations can be written
as
$t_{\sigma}^{(')}=\sum_{\sigma'}T_{\sigma\sigma'}^{(')}i_{\sigma'}^{(')}$,
$r_{\sigma}^{(')}=\sum_{\sigma'}R_{\sigma\sigma'}^{(')}i_{\sigma'}^{(')}$.
A careful examination shows that no spin flip amplitudes for transmission or
reflection in this spinor basis are present, and a possible modification of the
spinor is only due to a difference between propagating channels. Thus,
the scattering matrix reads 
\begin{equation}
\underline{S}=\left(\begin{array}{rrrr}
R_{\uparrow}&0&T_{\uparrow}'&0\\
0&R_{\downarrow}&0&T_{\downarrow}'\\
T_{\uparrow}&0&R_{\uparrow}'&0\\
0&T_{\downarrow}&0&R_{\downarrow}'\\
\end{array}\right).\label{eq:smat}
\end{equation}
The overall conductance then follows from the entries of the
scattering matrix by means of the Landauer conductance formula 
\cite{b"uttiker:85,landauer:87} shown in Eq.\eqref{eq:land}, with the spin
dependent transmission amplitude given by Eq.\eqref{eq:trm}.
The corresponding expression for the reflection amplitude is 
\begin{align}
R_{\sigma}(\phi,\beta,\gamma)&=&\frac{\cos\phi\gamma+3
\cos\phi-4\cos\Phi_{AC}^{\sigma}}{\cos\phi\gamma-5\cos\phi+4\cos\Phi_{AC}^{\sigma}+4i\sin\phi}.\nonumber\\ 
\end{align}
The (time reversed) functions for incident wave in the right lead are related to those above by
$T'_{\sigma}(\Phi_{AC}^{\sigma})=T_{\sigma}(-\Phi_{AC}^{\sigma})$ and $R'_{\sigma}=R_{\sigma}$.
 
The transmission and reflection coefficients with respect to the standard
$\sigma_{z}$-basis $\{|s\rangle\}$ are obtained by the corresponding spin rotation
$\Lambda (\{|s\rangle\}\rightarrow\{|\sigma\rangle\})$ of the
diagonal transmission and reflection blocks in the scattering matrix
\eqref{eq:smat}, e.g. for the transmission
\begin{equation}
T_{ss'}=\langle s'|\Lambda^{-1}(0)\circ
\big[T_{\sigma\sigma'}\big]\circ\Lambda(\varphi_{A})|s\rangle,
\end{equation}
where
\begin{equation}
\Lambda(\varphi)=\left(\begin{array}{cc}
\cos\frac{\theta}{2}&e^{-i\varphi}\sin\frac{\theta}{2}\\
\sin\frac{\theta}{2}&-e^{-i\varphi}\cos\frac{\theta}{2}
\end{array}\right),
\end{equation} and
\begin{equation}
\big[T_{\sigma\sigma'}\big]=\left(\begin{array}{cc}
T_{\uparrow}&0\\
0&T_{\downarrow}
\end{array}\right),
\end{equation}
and analogously for the reflection coefficients.

\bibliography{paper}

\end{document}